\documentstyle[aps,prb]{revtex} 

\begin{document}

\twocolumn[\hsize\textwidth\columnwidth\hsize\csname
@twocolumnfalse\endcsname

\title{The role of occupied $d$ states in the relaxation of hot electrons in
Au}
\author{I. Campillo$^{1}$, J. M. Pitarke$^{1,4}$,
A. Rubio$^{2}$, 
and P. M. Echenique$^{3,4}$}
\address{$^1$ Materia Kondentsatuaren Fisika Saila, Zientzi Fakultatea, 
Euskal Herriko Unibertsitatea,\\ 644 Posta kutxatila, 48080 Bilbo, Basque 
Country,
Spain\\
$^2$Departamento de F{\'{\i}}sica Te\'orica, Universidad de Valladolid,
47011 Valladolid, Spain\\
$^3$ Materialen Fisika Saila, Kimika Fakultatea, Euskal Herriko
Unibertsitatea,\\
1072 Posta kutxatila, 20080 Donostia, Basque Country, Spain\\
$^4$Donostia International Physics Center (DIPC) and Centro Mixto
CSIC-UPV/EHU,\\
Donostia, Basque Country, Spain}

\date{30 December 1999}

\maketitle

\begin{abstract}
We present first-principles calculations of electron-electron scattering rates of
low-energy electrons in Au. Our full band-structure calculations indicate that a
major contribution from occupied $d$ states participating in the screening of
electron-electron interactions yields lifetimes of electrons in Au with energies
of $1.0-3.0\,{\rm eV}$ above the Fermi level that are larger than those of
electrons in a free-electron gas by a factor of $\sim 4.5$. This prediction is
in agreement with a recent experimental study of ultrafast electron dynamics in
Au(111) films (J. Cao {\it et al}\,, Phys. Rev. B {\bf 58}, 10948 (1998)),
where electron transport has been shown to play a minor role in the measured
lifetimes of hot electrons in this material. 
\end{abstract}

]

\narrowtext

Relaxation lifetimes of excited electrons in solids with energies below the
vacuum level can be attributed to a variety of inelastic and elastic scattering
mechanisms, such as electron-electron (e-e), electron-phonon (e-p), and
electron-imperfection interactions.\cite{Petek3,review} Besides, when these
so-called hot electrons are generated by absorption of an optical pulse, as
occurs in the case of time-resolved two-photon photoemission (TR-2PPE)
techniques,\cite{Fauster,Haight} electron transport provides an additional decay
component to the photoexcited electron population. Since inelastic lifetimes of
hot electrons become infinitely long as they approach the Fermi level,
e-p scattering and the scattering by defects both play a key role in
the relaxation process of electrons very near the Fermi level. However, in the
case of hot electrons with energies larger than 
$\sim 0.5-1.0\,{\rm eV}$ above the Fermi level, e-e interactions yield
inelastic lifetimes that are in the femtosecond time scale and they provide the
main scattering mechanism.

Experimental femtosecond time-resolved photoemission studies of electron
dynamics have been performed in a variety of solid
surfaces,\cite{Fann,Schmu,Hertel,aes,Petek0,Cao1,Aes1,Knoesel1,Goldm,Aes98} the
role of e-e inelastic scattering and that of electron transport being difficult
to identify. However, recent TR-TPPE experiments in Au(111) films with
thicknesses ranging from $150$ to
$3000\,{\rm \AA}$ have shown the relaxation from electron transport to be
negligible and the hot-electron lifetime to be solely determined, at energies
larger than
$\sim 0.5-1.0\,{\rm eV}$ above the Fermi level, by e-e inelastic scattering
processes.\cite{Cao2} Hence, these measurements provide an excellent bench mark
against which to investigate the importance of band-structure and many-body
effects on electron dynamics in solids. Also, ballistic electron emission
spectroscopy (BEES) has shown to be capable of determining hot-electron
relaxation times in solid materials.\cite{BEEM}

In this paper, we report first-principles calculations of
the energy-dependent inelastic lifetime of hot electrons in Au. We follow
the many-body scheme first developed by Quinn and Ferrell\cite{QF} and by
Ritchie,\cite{Ritchie59} but we now include the full band structure of the
solid. This approach has already been successfully incorporated in the description
of inelastic lifetimes of excited electrons in a variety of simple (Al, Mg, and
Be) and noble (Cu) metals.\cite{Igorprl,Igorprb} A similar methodology, which is
also based on the GW approximation of many-body theory,\cite{Hedin} has been used
by other authors to evaluate hot-electron lifetimes in Al and Cu.\cite{Ekardt} 

Our many-body scheme for the calculation of inelastic lifetimes in
real solids has been described elsewhere.\cite{Igorprl,Igorprb} For periodic
crystals, the basic equation for the decay rate $\tau_i^{-1}$ of an electron in
the state $\phi_{{\bf k},n_i}$ of energy $E_{{\bf k},n_i}$ is (we use atomic
units throughout, i.e., $e^2=\hbar=m_e=1$)
\begin{eqnarray}\label{eq3}
\tau_i^{-1}={1\over \pi^2}\sum_{n_f}
\int_{\rm BZ}{{\rm d}{\bf q}}\sum_{{\bf G},{\bf G}'}&&
{B_{if}({\bf q}+{\bf G})B_{if}^*({\bf q}+{\bf G}')\over
\left|{\bf q}+{\bf G}\right|^2}\cr\cr
&&\times{\rm Im}\left[-\epsilon_{{\bf G},{\bf G}'}^{-1}({\bf
q},\omega)\right],
\end{eqnarray}
where $B_{if}({\bf q+G})$ represent the coupling between the hot-electron
initial and final states [see Eq. (5) of
Ref.\onlinecite{Igorprl}], and $\epsilon_{{\bf G},{\bf G}'}^{-1}({\bf q},\omega)$
are the Fourier coefficients of the inverse dielectric function of the solid,
which we compute in the random-phase approximation (RPA).\cite{Fetter} ${\bf G}$
and
${\bf G}'$ represent reciprocal lattice vectors, the integration over ${\bf q}$
is extended over the first Brillouin zone (BZ), the sum over
$n_f$ is extended over the band structure for each wave vector in the first
BZ, $\omega=E_{{\bf k},n_i}-E_{{\bf k}-{\bf q},n_f}$ is the energy transfer
($0<\omega<E_{{\bf k},n_i}-E_F$), and $E_F$ is the Fermi energy.

If one neglects crystalline local-field effects, Eq. (\ref{eq3}) can be
expressed as follows
\begin{equation}\label{eq17}
\tau_i^{-1}={1\over\pi^2}\sum_{n_f}
\int_{\rm BZ}{{\rm d}{\bf q}}\sum_{\bf G}
{\left|B_{if}({\bf q}+{\bf G})\right|^2\over\left|{\bf q}+{\bf
G}\right|^2}
{{\rm Im}\left[\epsilon_{{\bf G},{\bf G}}({\bf q},\omega)\right]\over
|\epsilon_{{\bf G},{\bf G}}({\bf q},\omega)|^2}.
\end{equation}
Here, initial and final states of the hot electron enter through the
coefficients $B_{if}({\bf q}+{\bf G})$. The imaginary part of $\epsilon_{{\bf
G},{\bf G}}({\bf q},\omega)$ represents a measure of the number of states
available for the creation of an electron-hole pair involving a given
momentum and energy transfer ${\bf q}+{\bf G}$ and $\omega$,
respectively, renormalized by the coupling between electron and hole states.
The factor $\left|\epsilon_{{\bf G},{\bf G}}({\bf q},\omega)\right|^{-2}$
accounts for the dynamically screened e-e interaction.

The decay rate of Eqs. (\ref{eq3}) and (\ref{eq17}) depends on both the wave
vector ${\bf k}$ and the band index $n_i$ of the initial Bloch state. Since
measurements of hot-electron lifetimes are reported as a function of
energy and the proper choice of the wave vector ${\bf k}$ and the band index
$n_i$ is usually not easy to be determined, we define $\tau^{-1}(E)$ as the
average of $\tau^{-1}({\bf k},n)$ over all wave vectors and bands lying with
the same energy in the irreducible wedge of the Brillouin zone (IBZ).  

If all one-electron Bloch states entering both the coefficients
$B_{if}({\bf q}+{\bf G})$ and the dielectric function $\epsilon_{{\bf G},{\bf
G}'}({\bf q},\omega)$ were
represented by plane waves and, at the same time, all energy bands were
replaced by those of free electrons, then decay rates would not be
direction-dependent and both Eqs. (\ref{eq3}) and (\ref{eq17}) would exactly
coincide with the GW scattering rate of hot electrons in a FEG.\cite{review} For
hot electrons with energies very near the Fermi level ($E\sim E_F$) this result
leads, in the high-density limit ($r_s<<1$),\cite{rs} to the well-known formula
of Quinn and Ferrell\cite{QF} [see Eq. (7) of Ref.\onlinecite{Igorprl}, which
yields time in {\rm fs} when $E$ and $E_F$ are expressed in ${\rm eV}$]. The
$r_s^{-5/2}$ scaling of hot-electron lifetimes predicted by this equation is
found to be entirely originated in the density-dependent screening of e-e
interactions [high electron densities yield a weaker interaction, $\sim
r_s^{-3/2}$] and the density-dependent momentum of the hot electron [high
electron densities yield a larger momentum and, therefore, a smaller number of
available transitions, $\sim r_s^{-1}$].

Gold is a noble metal with entirely filled $5d$-like bands. In Fig. 1 we show
the energy bands of this face-centered cubic crystal. We see large
differences between the band structure of this material and that of free
electrons. Slightly below the Fermi level, at $E-E_F\sim 2\,{\rm eV}$, we have
$d$ bands capable of holding $10$ electrons per atom, the one remaining
electron being in a free-electron-like band below and above the $d$ bands.
These $d$ bands, which are concentrated within a width of $\sim 5.5\,{\rm eV}$,
are associated with the characteristic $d$-band wave functions at levels
$\Gamma_{25}'$ and $\Gamma_{12}$. Hence, a combined description of both
delocalized $6s^1$ and localized $5d^{10}$ electrons is needed in order to
address the actual electronic response of Au. The results presented below have
been obtained by first expanding all one-electron Bloch states in a plane-wave
basis and then solving the Kohn-Sham equation\cite{Kohn} of density-functional
theory (DFT)\cite{XC} with all $6s^1$ and $5d^{10}$ Bloch states taken as valence
electrons in the pseudopotential generation.\cite{note} Thus, a kinetic-energy
cutoff as large as
$75\,{\rm Ry}$ has been required, thereby keeping $\sim 1200$ plane waves in
the expansion of each Bloch state. Though all-electron schemes,
such as the full-potential linearized augmented plane-wave (LAPW)
method,\cite{LAPW} are expected to be better suited for the description of the
response of localized $d$ electrons, the plane-wave pseudopotential approach
has already been successfully incorporated in the description of the dynamical
response\cite{Igorcu} and hot-electron lifetimes\cite{Igorprl,Igorprb,Ekardt} of
Cu. 

The sampling over the BZ required for the evaluation of both the dielectric
matrix and the hot-electron decay rate of Eqs. (\ref{eq3}) and (\ref{eq17}) has
been performed on a $16\times 16\times 16$ Monkhorst-Pack (MP) mesh.\cite{Monk}
For hot-electron energies under study ($E-E_F\sim 1.0-3.0\,{\rm eV}$),
well-converged results have been found with the inclusion in the evaluation of
the dielectric matrix of conduction bands up to a maximum energy of $25\,{\rm
eV}$ above the Fermi level. The sums in Eqs. (\ref{eq3}) and (\ref{eq17}) have
been extended over $15\,{\bf G}$ vectors, the magnitude of the maximum momentum
transfer ${\bf q}+{\bf G}$ being  well over the upper limit of $\sim 2\,q_F$
($q_F$ is the Fermi momentum).  

Our full band-structure calculation of the average lifetime $\tau(E)$ of hot
electrons in Au, as obtained from Eq. (\ref{eq3}) with full inclusion of
crystalline local-field effects, is presented in Fig. 2 by solid circles. The
lifetime of hot electrons in a FEG with the electron density equal to that of
$6s^1$ electrons in Au ($r_s=3.01$) is exhibited in the same figure, as obtained
from either Eq. (\ref{eq3}) or Eq. (\ref{eq17}) with the full RPA dielectric
function of free electrons (solid line) and from the formula of Quinn and
Ferrell (dotted line). Also plotted in this figure is the lifetime of
hot electrons in Au(111) films, as determined from accurate TR-TPPE
experiments (open squares).\cite{Cao2} The agreement between our full
band-structure calculation and the experimental data is excellent, for all
hot-electron energies under study.

Both band-structure (solid circles) and FEG (solid line) calculations presented
in Fig. 2 have been carried out within the same many-body framework: The
electron self-energy has been obtained in the GW approximation and {\it
on-the-energy-shell} [deviations of the actual excitation energy from the
independent-particle approximation have been neglected], and the RPA has been
used to compute the dielectric function of the solid. Hence, the ratio ($\sim
4.5$) between our calculated {\it ab initio} lifetimes and the corresponding
FEG calculations unambiguously establishes the impact of the band structure of
the crystal on the hot-electron lifetime.\cite{note3} Furthermore, since
many-body effects beyond our GW-RPA scheme are expected to be within $20\%$ of
our calculations,\cite{note1} we conclude that large deviations of the
experimental measurements from the FEG prediction are mainly
due to band-structure effects. Nevertheless, one must be cautious
with the comparison between our full band-structure calculations and the
experiment. Since Au has a wide energy gap along the $\Gamma L$ direction just
over the Fermi level, the $k_\parallel=0$ photoemission in Au(111) cannot be
associated, in the energy range under study ($E-E_F\sim 1.0-3.0\,{\rm eV}$),
with hot-electron Bloch states in the same direction, and therefore the proper
choice of the wave vector ${\bf k}$ appears to be unclear. We have evaluated
hot-electron lifetimes along various directions of the wave vector, and have
found that differences between these results and the average lifetimes presented
in Fig. 2 are within $\sim 20\%$, which gives support to our comparison with the
experiment.\cite{note2}

In order to investigate the role that occupied $d$ states plays in the relaxation
of hot electrons in Au, now we neglect crystalline local-field effects and
compute hot-electron lifetimes from Eq. (\ref{eq17}). Scaled lifetimes
$\tau(E)\times(E-E_F)^2$ of hot electrons in Au, as obtained from Eq.
(\ref{eq17}) by replacing the hot-electron Bloch states and bands entering
$|B_{if}({\bf q}+{\bf G})|^2$ by those of free electrons and the dielectric
function in $|\epsilon_{{\bf G},{\bf G}}({\bf q},\omega)|^{-2}$ by that of a FEG
with $r_s=3.01$, are represented in the inset of Fig. 2 by open triangles. These
calculated lifetimes, which have been obtained with full inclusion of the band
structure of the crystal in the evaluation of ${\rm Im}[\epsilon_{{\bf G},{\bf
G}}({\bf q},\omega)]$,  are close to those obtained within the FEG model of the
solid (solid line), showing that the combined effect of the density of states
(DOS) available for e-h pair creation and the small overlap between $d$
states below and $sp$ states above the Fermi level nearly compensate. Hence,
deviations of actual hot-electron lifetimes from FEG predictions are mainly
originated in the deviation of the hot-electron momentum from the FEG prediction
[the momentum of hot electrons depends on the actual DOS at the Fermi level] and
the participation of $d$ electrons in the screening of e-e interactions.

The effect of virtual interband transitions giving rise to
additional screening is to largely increase the hot-electron lifetime in Au, as
occurs in the case of Cu,\cite{Igorprl} a noble metal with analogous
electronic structure. Since the change $\delta\epsilon_1^b$ in
$\epsilon_1(\omega)$ due to the presence of $d$ electrons in the noble metals is
found to be practically constant at low frequencies,\cite{eh1,eh2}
Quinn\cite{Quinn63} treated the FEG as if it were embedded in a medium of
dielectric constant $\epsilon_0=1+\delta\epsilon_1^b$ instead of unity. The
corrected lifetime is then found to be larger by roughly a factor of
$\epsilon_0^{1/2}$, i.e., $\sim 2.5$ for both Cu and Au. Nevertheless, our
first-principles evaluation of the wave-vector and frequency
dependent dielectric matrix leads us to the conclusion that the role that
occupied $d$ states play in the screening of e-e interactions is much more
important in Au than in Cu. This is an expected result, since Au $5d$ electrons
lie further away from the nucleus than Cu $3d$ electrons, and $5d$ bands in Au
are, therefore, more free-electron-like than $3d$ bands in Cu [$d$ bands in Cu
are concentrated within a width of $\sim 3\,{\rm eV}$,\cite{Burdick} smaller
than in the case of Au]. 

The result we obtain with full inclusion of the band structure of the crystal in
the evaluation of Eq. (\ref{eq17}), but still neglecting crystalline local-field
effects, is represented in the inset of Fig. 2 by open circles. Our full
band-structure calculation of Eq. (\ref{eq3}) for Au is represented in the same
figure by solid circles, showing that neglecting crystalline local-field
corrections results in an overestimation of hot-electron lifetimes of
$40\%-50\%$. These crystalline local-field effects, which partially compensate
the pronounced participation of
$d$ electrons in the screened e-e interaction, are due to large electron-density
variations in this material.

In conclusion, we have presented full band-structure calculations of the
inelastic lifetime of hot electrons in Au. We have found that a major
contribution from occupied $d$ states participating in the screening of e-e
interactions yields lifetimes of electrons with energies of
$1.0-3.0\,{\rm eV}$ above the  Fermi level that are larger than those of
electrons in a FEG by a factor of $\sim 4.5$, in agreement with the
experiment. The effect of virtual interband transitions giving rise to
additional screening does not depend on whether the hot electron can excite
$d$ electrons [the $d$-band scattering channel opens at $\sim 2\,{\rm eV}$
below the Fermi level] or not. On the other hand, the actual density of
occupied states available for real transitions is found not to play an
important role in the relaxation mechanism. Consequently, actual lifetimes
approximately scale as $(E-E_F)^{-2}$, as in the case of a FEG. That lifetimes
of hot electrons in Au can be fitted within this scaling behaviour has also
been demonstrated by a theoretical analysis of electron-electron mean free paths
in BEES experiments.\cite{BEEM}

We acknowledge partial support by the University of the Basque Country, the
Basque Hezkuntza, Unibertistate eta Ikerketa Saila, and the Spanish Ministerio de
Educaci\'on y Cultura.

\begin{figure}
\caption[]{Calculated band structure of Au along certain symmetry directions.}
\end{figure}

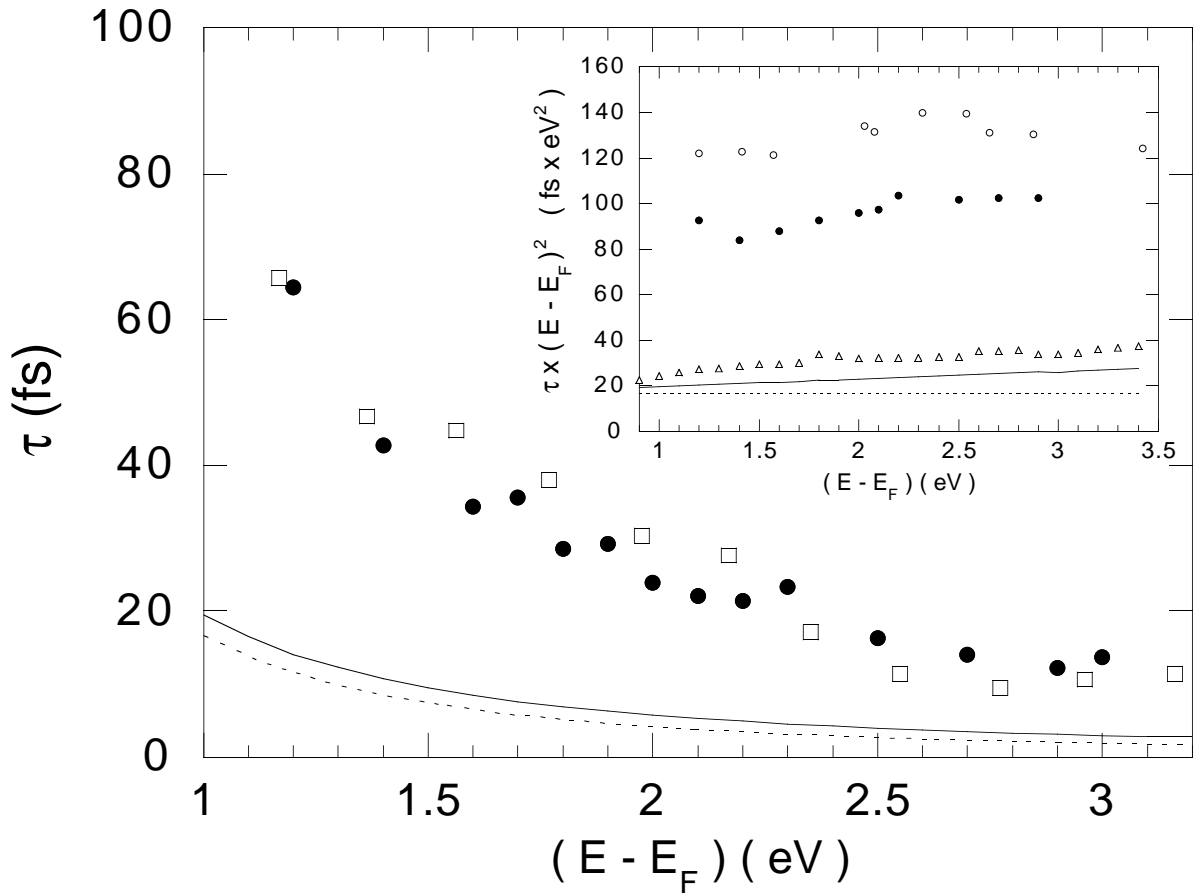
\begin{figure}
\caption[]{Hot-electron lifetimes in Au. Solid circles represent  our full {\em
ab initio} calculation of $\tau(E)$, as obtained  after averaging $\tau({\bf
k},n)^{-1}$ of Eq. (\ref{eq3}) over wave vectors and over the band structure for
each ${\bf k}$. The solid line represents the lifetime of hot electrons in a FEG
with $r_s=3.01$, as obtained within the full GW-RPA. Open squares represent the
experimental measurements of Ref.\onlinecite{Cao2}. The dotted line represents
the prediction of Quinn and Ferrell. Differences between the FEG calculations
represented by solid and dotted lines are due to the fact that the formula of
Quinn and Ferrell is obtained for hot-electron energies very near the Fermi level
($E\sim E_F$) and in the high-density limit ($r_s\to 0$). The inset exhibits
scaled lifetimes of hot electrons in Au. Open circles represent our calculation
of
$\tau(E)$, as obtained  after averaging $\tau({\bf k},n)^{-1}$ of either Eq.
(\ref{eq3}) or Eq. (\ref{eq17}) over wave vectors and over the band structure
for each
${\bf k}$ and neglecting crystalline local-field corrections. Open
triangles represent the result obtained from Eq. (\ref{eq17}) by replacing
hot-electron Bloch states in $B_{if}({\bf q}+{\bf G})$ and $|\epsilon_{{\bf
G},{\bf G}}({\bf q},\omega)|^{-2}$ by those of free electrons, but with full
inclusion of the band structure of the crystal in the evaluation of ${\rm
Im}[\epsilon_{{\bf G},{\bf G}}({\bf q},\omega)]$.}
\end{figure} 


\begin{references}
\bibitem{rs} The so-called electron-density parameter $r_s$ is defined by
the relation $1/n_0=(4/3)\pi r_s^3$, $n_0$ being the average electron
density.
\bibitem{note2} The anisotropy of hot-electron lifetimes in Au is only found to
play an important role at higher energies, where Bloch states with the wave
vector along the necks of the Fermi surface, in the $\Gamma L$ direction, are
available.
\end{references}
\end{document}